\documentclass[a4paper,11pt]{article}

\usepackage{amsmath}
\usepackage{a4wide}
\usepackage{amsfonts}
\newenvironment{pf}{\textbf{Proof:}}{\hspace{\stretch{1}}$\square$}

\newtheorem{thm}{Theorem}

\newtheorem{pr}[thm]{Proposition}

\newcommand{\rH}{\mathcal{H}}

\font\timesept=cmr7

\newcommand{\NN}{\mathbb{N}}

\newcommand{\RR}{\mathbb{R}}

\def\Rp{\RR^+}
\def\tr{\mathop{\rm Tr}\nolimits}

\title{Stochastic Master Equations\\ in Thermal Environment\footnote{Work supported by ANR project ``HAM-MARK" N${}^\circ$ ANR-09-BLAN-0098-01.}}
\author{St\'ephane $\textrm{ATTAL}^1$ and Cl\'ement $\textrm{PELLEGRINI}^2$}

\date{}
\begin{document}
\maketitle
\vskip -0.5cm
\centerline{\timesept $^1$ Universit\'e de Lyon, Universit\'e de Lyon 1}
\vskip -1mm
\centerline{\timesept Institut Camille Jordan, U.M.R. 5208}
\vskip -1mm
\centerline{\timesept 21 av Claude Bernard}
\vskip -1mm
\centerline{\timesept 69622 Villeurbanne cedex, France}
\centerline{\timesept $^2$ Institut de Math\'ematiques de Toulouse }
\vskip -1mm
\centerline{\timesept Laboratoire de Statistique et de Probabilit\'e}
\vskip -1mm
\centerline{\timesept Universit\'e Paul Sabatier (Toulouse III)}
\vskip -1mm
\centerline{\timesept 31062 Toulouse Cedex 9, France}

\begin{abstract}
We derive the stochastic master equations which describe the evolution of open quantum systems in contact with a heat bath and undergoing indirect measurements. These equations are obtained as a limit of a quantum repeated measurement model where we consider a small system in contact with an infinite chain at positive temperature. At zero temperature it is well-known that one obtains stochastic differential equations of jump-diffusion type. At strictly positive temperature, we show that only pure diffusion type equations are  relevant. 
\end{abstract}

\section{Introduction}

The theory of \textit{Open Quantum Systems} aims to study the time evolution of a small system $\mathcal{H}_0$ interacting with an
environment $\mathcal{E}$, cf \cite{SA,c,francesco2,carm}. Starting from an \textit{Hamiltonian} description of the coupled system \cite{c,francesco2,carm}, the evolution of the reduced
system $\mathcal{H}_0$ is obtained by tracing over the degree of freedom of the environment. 

In the \textit{Markovian} approach of
open systems, the time evolution of the state of the reduced system is characterized by a semigroup of completely positive maps, with a typical generator called
\textit{Lindblad generator}, which gives rise to an ordinary differential equation called \emph{master equation} \cite{c,francesco2,gardiner}. 

In this framework, an
active line of research, motivated by recent experimental applications in quantum optics and quantum communications, is
focused on the description of quantum measurement \cite{Book,MR1121822,MR1919502,MR1639788,barchielli0,barchielli1,belav1,belav3,francesco2,haroche,gardiner,carm}. Basically, in order to avoid \textit{Zeno effect} \cite{francesco2}, the measurement is
performed on the environment. According to the postulates of quantum mechanics, this involves a random perturbation of the evolution of
the state of $\mathcal{H}_0$. The dynamics of $\mathcal{H}_0$ is then described by classical stochastic differential equations, which
are perturbations of the master equation in terms of white noise \cite{Book,b1,b2,MR1121822,MR1919502,MR1639788,barchielli0,barchielli1,SSEMB,infinite,P1,P2,P3}. Usually, these equations are called \textit{Stochastic
Schr\"odinger Equations} or \textit{Stochastic Master Equations} and their solutions are called \textit{Quantum Trajectories} (the
name ``stochastic Schr\"odinger equation" is usually reserved for the evolution of the state of $\mathcal{H}_0$ in terms of pure states
whereas stochastic master equations concerns evolution of density matrices). 

In the literature, most of the results concern models
where the environment is at zero temperature and a lot of investments are still in progress to get right descriptions for model at
positive temperature (even without measurement). Typically, at zero temperature, for the random evolution of density matrices we get two typical stochastic master equations of the following form.

\begin{enumerate}
\item A diffusive equation
\begin{equation}\label{D} d\rho_t=\mathcal{L}(\rho_t)\,dt+\Big(C\rho_t+\rho_tC^\star-\tr\Big[\rho_t(C+C^\star)\Big]\,\rho_t\Big)\,dW_t,
\end{equation}
where $W_t$ is a one dimensional Brownian motion.
\item A jump equation
\begin{equation}\label{J}
d\rho_t=\mathcal{L}(\rho_t)\, dt+\left(\frac{C\,\rho_t\,C^\star}{\tr\big[C\,\rho_t\,C^\star\big]}-\rho_t\right)\,\Big(d\tilde{N}_t-\tr\big[C\,\rho_t\,C^\star\big]\,dt\Big),
\end{equation}
where $(\tilde{N}_t)$ is a counting process with stochastic intensity $t\rightarrow\int_0^t\tr\big[C\,\rho_s\,C^\star\big]ds.$
\end{enumerate}
In the above expressions $\mathcal{L}$ is the Lindblad operator. Note that we can recognize from the form of these equations that they all are \emph{simulations} of the master equation, for they are stochastic differential equations valued in the set of states of $\rH_0$ and on average they satisfy the master equation.

 More complicated models use jump-diffusion stochastic differential
equations which are mixing of equations (\ref{D}) and (\ref{J}) \cite{MR1919502,barchielli0,P3} (see Section \ref{discretemodel}).

\smallskip
Mathematically, there are three usual ways to justify these equations. A first approach is based on \textit{Instrumental Operator
Processes} connected with the notion of \textit{Operator Valued Measures} and \textit{Quantum Markov Semi-groups} analogy
\cite{c,Book}. The second one is based on classical stochastic differential equation theory and on the concept of \textit{a posteriori
state} \cite{Book,MR1121822,MR1919502,MR1639788,barchielli0,infinite}. The third approach often called \textit{Quantum Filtering} is based on the formalism of \textit{Quantum Stochastic Calculus}
and the notion of input and output field in quantum optics \cite{c,barchielli1,belav1,belav3,SSEMB,b1,b2}. In this last setup, the evolution of the small system and the environment
is modeled by \textit{Quantum Langevin Equations} (also called \textit{Quantum Stochastic Differential Equations} or
\textit{Hudson-Parthasarathy Equations}) \cite{SA,c,PS}. These equations are namely driven by \textit{Quantum Noises} \cite{SA,c}. Next, by
adapting the classical framework of stochastic filtering, one can obtain appropriate stochastic differential equations driven by
classical noises, which take into account the ``incomplete" information of indirect observations.

In \cite{P1,P2,P3}, an alternative discrete way has been developed.  The approach is based on the model of \textit{Quantum Repeated
Interactions} which provides a ``useful" discrete approximation model of quantum Langevin equations \cite{FLA,FRTC}. The setup of quantum
repeated interactions is the one of the interaction of a small system $\mathcal{H}_0$ with an environment represented by an infinite
chain $\bigotimes_{k=0}^\infty\mathcal{H}_k$. Moreover, the pieces of the chain are identical and independent quantum system, that is
$\mathcal{H}_k=\mathcal{H}$ for all $k$. Each copy $\mathcal{H}$ interacts with $\mathcal{H}_0$, one after the other, during a time
$\tau$. In this framework, an appropriate language of \textit{discrete quantum noise} and \textit{discrete quantum stochastic
differential equations} is constructed. Next, it is shown that the time continuous limit ($\tau\rightarrow0$) gives rise to continuous
models of quantum Langevin equations and continuous quantum noises.  In this context, a discrete time model of indirect quantum
measurements, called model of \textit{Quantum Repeated Measurements} has been developed in \cite{P1,P2,P3}.  It consists in performing a
measurement of an observable of  $\mathcal{H}_k$ after each interaction with $\mathcal{H}_0$. As in the continuous case, the
measurement introduces a random perturbation of the evolution of $\mathcal{H}_0$, described by \textit{discrete stochastic master
equations} and \textit{discrete quantum trajectories}. In the same spirit as \cite{FRTC}, by considering the continuous time limit
($\tau\rightarrow0$), the continuous models of stochastic master equations and quantum trajectories are recovered. Much beyond the
approximation result, these discrete time models (without measurement \cite{FRTC} and with measurement \cite{P1,P2,P3}) provide a concrete and
intuitive physical justification of the continuous models of quantum Langevin and stochastic master equations.

\bigskip
It is important to notice that these four different approaches of stochastic master equations are crucially based on the ``zero
temperature" assumption. For example in the approach based on quantum stochastic calculus, one the main obstacle to consider positive
temperature model is to describe the action of the heat bath on the small system and to derive adequate Langevin equations.

 Recently, in \cite{MR2323437}, the discrete approach of quantum repeated interactions \cite{FRTC}
 has been adapted to models with heat bath at positive temperature. This way, thermal quantum
 Langevin equations have been obtained and a clear justification of the action of the thermal bath has been presented.
 It is then natural to combine this approach with quantum repeated measurements in order to derive stochastic master
 equations for heat baths. 
 
 In order to introduce temperature, we consider that each copy $\mathcal{H}$ is in a thermal
 Gibbs state at inverse temperature $\beta$. The crucial point in all the different approaches at zero temperature
 is the fact the state of the environment is a pure state. This is clearly not the case for a Gibbs state. In order to
 get around this difficulty, we apply the G.N.S. representation of that state. This way, the Gibbs state of each copy
 $\mathcal{H}$ can be considered as a pure state in an enlarged Hilbert space. Hence, with this representation,
 the convergence result of \cite{P3} can be applied and stochastic master equations for heat bath are derived. Surprisingly, models of the form (\ref{J}) with counting processes disappears and only diffusive models remains.

\bigskip
This article is structured as follows.

In Section \ref{discretemodel}, we remind the discrete models of quantum repeated interactions and then quantum repeated measurements. Next, we recall the main result of \cite{P3} which gives the stochastic Master equations as continuous limits of these discrete models.

In Section \ref{c}, we adapt the result of \cite{P3} for model with positive temperature. This is achieved by describing the G.N.S
representation of the heat bath. Hence we obtain the complete description of stochastic master equations for a small system in contact
with a heat bath and undergoing indirect quantum measurements.

\section{From Discrete to Continuous Quantum Trajectories at Zero Temperature}\label{discretemodel}

In this section we recall the mathematical description of the quantum repeated measurement model, as developed in \cite{P3}.

\subsection{Quantum Repeated Interactions}\label{QRI}

The model of quantum repeated interactions consists in studying the interaction of a finite dimensional quantum system $\mathcal{H}_0$ in contact with an infinite chain $T\Phi=\bigotimes_{\mathbb{N}^\star}\mathcal{H}_k$, where $\mathcal{H}_k=\mathcal{H}=\mathbb{C}^{N+1}$ for all $k$.

We first need to make precise the definition of the countable tensor product $T\Phi$. To this end, let us define an explicit orthonormal basis of $T\Phi$. Let $\{X_0,X_1,\ldots,X_N\}$ be an orthonormal basis of $\mathcal{H}=\mathbb{C}^{N+1}$. Actually note that only the choice of $X_0$ is relevant in our construction: it represents a reference state of $\rH$ (for example a ground state). Put $X^n_i$ to be the copy of the basis vector $X_i$ but acting on the $n$-th copy of $\rH$.
The orthonormal basis of $T\Phi$ is then made of those tensor products
$$
X^1_{i_1}\otimes\ldots\otimes X^n_{i_n}\otimes\ldots
$$
such that all the $i_n$'s, but a finite number, are all equal to 0.

On $\rH$ consider the  basic operators $a_j^i, i,j=0,\ldots,N$ defined by  $a_j^iX_k=\delta_{ik}X_j$ (in Dirac notation $a_j^i=\vert X_j\rangle\langle X_i\vert$). We dilate them as operators $a_j^i(k)$ on $T\Phi$ by asking them to act as $a_j^i$ on the $k$-th copy of $\mathcal{H}$ and as the identity operator on the rest of the chain.

\smallskip
The model of repeated interactions \cite{FRTC} is now described as follows. Each copy of $\mathcal{H}$ is supposed to interact, one after the other, with $\mathcal{H}_0$ during a time duration $\tau$. Each elementary interaction between $\rH_0$ and $\rH$ is described by a total Hamiltonian 
\begin{equation}\label{Htot}
H_{\rm tot}=H_0\otimes I+I\otimes H_R+\lambda H_I\,.
\end{equation}
The operator $H_0$ corresponds to the free Hamiltonian of the system $\mathcal{H}_0$, the operator $H_R$ is the free Hamiltonian of the system $\mathcal{H}$, the operator $H_I$ is the interaction Hamiltonian and $\lambda$ is the coupling constant. 

The basis $\{X_0, \ldots, X_n\}$ is chosen to be the basis of eigenvectors of $H_R$, that is, with our notations:
\begin{equation}\label{HR}
H_R=\sum_{i=1}^N\gamma_i\,a_i^0a_0^i\,.
\end{equation}
The interaction Hamiltonian $H_I$ is chosen to be of so-called ``dipole-type":
\begin{equation}\label{HI}
H_I=\sum_{i=1}^N\big(C_i\otimes a_i^0+C_i^\star\otimes a_0^i\big)\,.
\end{equation}
After a time duration $\tau$ of interaction, the evolution of $\rH_0\otimes \rH$ is governed by the unitary operator
$$
U=e^{-i\tau H_{tot}}\,.
$$
That is, in the Schr\"odinger picture, the evolution of states on $\mathcal{H}_0\otimes\mathcal{H}$ is given by
$$
\rho\mapsto U\rho U^\star
$$
and in the Heisenberg picture, the observables evolve as
$$
X\mapsto U^\star\rho U\,.
$$
Now, in order to describe the repeated interactions on the whole space $\mathcal{H}_0\otimes T\Phi$, we consider, for each $k\in\NN^*$,  the unitary operator
$U_k$ which acts like the operator $U$ on the tensor product
$\mathcal{H}_0\otimes\mathcal{H}_k$ and like the identity operator on the
rest of the space. For $k$ being fixed, the operator $U_k$ describes the $k$-th interaction. The whole procedure is then described by the
sequence of unitary operators $(V_k)$ defined by $V_k=U_kU_{k-1}\ldots U_1$. For example the evolution of an initial state $\rho$ on $\rH_0\otimes T\Phi$ after $k$
interactions is given by
$$
\rho\mapsto V_k\rho V_k^\star\,.
$$

\subsection{Quantum Repeated Measurements}\label{QRM}

 Now, we are in position to describe the model of quantum repeated measurements \cite{P1}. To this end, we need to specify the reference states of $\mathcal{H}_0$ and $\mathcal{H}$. Let $\rho$ denote the initial state of $\mathcal{H}_0$. For each copy of $\mathcal{H}$, we consider the usual thermal Gibbs state at inverse temperature $\beta$:
\begin{equation}
\rho_\beta=\frac{e^{-\beta H_R}}{\tr\left[e^{-\beta H_R}\right]},
\end{equation}
where $H_R$ is defined in expression (\ref{HR}). In particular $\rho_\beta$ is diagonal, with diagonal elements that we shall denote by $\{\beta_0,\beta_1,\ldots,\beta_N\}$.

\smallskip
 In order to describe the indirect measurement of an observable $A$ of $\mathcal{H}$, we come back to the description of the first interaction in the space $\mathcal{H}_0\otimes\mathcal{H}$. After the interaction the new state of $\mathcal{H}_0\otimes\mathcal{H}$ is $\mu=U(\rho\otimes\rho_\beta)U^\star$. Now if $A$ owns the spectral decomposition
 $$A=\sum_{i=0}^p\lambda_iP_i\,,$$
the measurement of $A$ gives a random result in the set of eigenvalues $\lambda_0,\ldots, \lambda_p$. In particular, the value $\lambda_i$ is obtained with probability
\begin{equation}
P\big[\,\textrm{to observe}\,\,\lambda_i\,\big]=\tr\big[\mu\,I\otimes P_i\big]\,.
\end{equation}
After having observed the eigenvalue $\lambda_i$, the state $\mu$ is projected and becomes
$$
\tilde{\rho}_1(i)=\frac{(I\otimes P_i)\,\mu\,(I\otimes P_i)}{Tr\big[\mu\,(I\otimes P_i)\big]}\,.
$$
 For $i$ being fixed, the state $\tilde{\rho}_1(i)$ represents the new state of $\mathcal{H}_0\otimes\mathcal{H}$ after the first interaction and the first measurement. Usually, we are only interested in the reduced system $\mathcal{H}_0$, that is, we shall consider only the partial trace
$$\rho_1(i)=\tr_{\rH}\left(\tilde{\rho}_1(i)\right)\,.$$
The state $\rho_1$ is a random state which takes the values 
$$
\frac{\mathcal{L}_i(\rho)}{\tr\big[\mathcal{L}_i(\rho)]}
$$ 
with
probability $Tr\big[\mathcal{L}_i(\rho)]$ respectively, where 
$$\mathcal{L}_i(\rho)=\tr_{\rH}\big[(I\otimes
P_i)\,U(\rho\otimes\rho_\beta)U^\star\,(I\otimes P_i)\big]\,.
$$
This way, the state $\rho_1$ is the new reference state of
$\mathcal{H}_0$ and we can consider a new interaction with a copy of $\mathcal{H}$ and a new measurement of $A$. By reducing on
$\mathcal{H}_0$, we then get a new random state $\rho_2$ which satisfies similar properties as $\rho_1$. Iterating this procedure we obtain
a random sequence of state $(\rho_k)$ on $\rH_0$. This sequence, called \textit{discrete quantum trajectory} describes the random modifications of the state of $\mathcal{H}_0$ undergoing
quantum repeated interactions and measurements. More precisely, we have the following proposition (cf \cite{P1}).

\begin{pr}
The sequence $(\rho_k)$ is a Markov chain. More precisely, if $\rho_k=\theta$, the state $\rho_{k+1}$ can take the values
$$\frac{\mathcal{L}_i(\theta)}{\tr\big[\mathcal{L}_i(\theta)]},\,\,i=0,\ldots,p$$
with probability $p_i(\theta)=\tr\big[\mathcal{L}_i(\theta)]$, where $\mathcal{L}_i(\theta)=\tr_{\rH}\big[I\otimes P_i\,\,U(\theta\otimes\rho_\beta)U^\star\,\,I\otimes P_i\big]$.
\end{pr}

In other words, the above proposition can be summarized as follows. Let $\mathbf{1}_i^{k}$ denote the random variable which takes the value $1$ if we observe the eigenvalue $\lambda_i$ during the
$k$-th measurement and $0$ otherwise, we then have
\begin{equation}\label{discschro}
\rho_{k+1}=\sum_{i=0}^p\frac{\mathcal{L}_i(\rho_k)}{p_i(\rho_k)}\mathbf{1}_i^{k+1}\,.
\end{equation}

\subsection{Quantum Trajectories at Zero Temperature}\label{QTZT}

The equation above is a discrete-time stochastic master equation. In the articles \cite{P1,P2,P3} the author computes explicitely the continuous-time limit of (\ref{discschro}) at zero temperature. In the limit, he obtains the usual stochastic master equations describing continuous measurement experiments. In general, these equations are stochastic differential equations mixing diffusive and jump noises. We shall now recall these results.

\smallskip
We now focus on the case where the temperature is zero, that is $\beta=+\infty$. This way, we have $\beta_i=0$ for all $i=1,\ldots,N$ and $\rho_\beta=a_0^0=\vert X_0\rangle\langle X_0\vert$. 

If $K$ is an operator on $\rH_0\otimes\rH$ and for the choice $\{X_0, \ldots, X_N\}$ of an orthonormal basis for $\rH$, the operator $K$ can be written as a $(N+1)\times(N+1)$-block-matrix, with coefficients 
$$
K^i_j=\tr_{\rH}\Big[\big(I\otimes\vert X_j\rangle\langle X_i\vert\big)\, K\Big]\,,
$$
being operators on $\rH_0$. With these notations, we have 
$$
\tr_{\rH}[K]=\sum_{i=0}^N K_i^i\,.
$$ 
In particular the $\mathcal{L}_i(\rho)$ are easy to compute explicitely: let $U=(U_k^l)_{0\leq
k,l\leq N}$ be the block-matrix representation of the unitary evolution $U$ and let $P_i=(p_{kl}^i)_{0\leq k,l\leq N}$ in the basis $\{X_0,\ldots,X_1\}$ (in block form we have $I\otimes P_i=(p_{kl}^i\,I)_{0\leq k,l\leq N}$), we get
\begin{equation}\label{exprLi}
\mathcal{L}_i(\rho)=\tr_{\rH}\big[I\otimes P_i\,\,U(\rho\otimes\rho_\beta)U^\star\,\,I\otimes P_i\big]=\sum_{k,l=0}^Np_{kl}^i\,\,U_k^0\,\rho\,(U_l^0)^\star\,.
\end{equation}

The convergence result of discrete quantum trajectories is based on the asymptotic assumptions described in \cite{FRTC}. These assumptions concern the unitary operator $U$. If we consider the time of interaction $\tau$ being $\tau=1/n$, the unitary operator $U$ depends on the parameter $n$, that is, $U=U(n)=(U_k^l(n))_{0\leq k,l\leq N}$. In \cite{FRTC}, it is shown that the operator process $(V_{[nt]})$ satisfying
$$V_{[nt]}=U_{[nt]}(n)\ldots U_{1}(n)$$
converges, non trivially, to a process $(V_t)$, only if the coefficients $U_j^i(n)$ obey certain normalizations. The limit process $(V_t)$ then satisfies a quantum Langevin equation describing the evolution of a small system coupled with a Fock space. 

Their asymptotic conditions concern the existence of operators $L_j^i$ such that for all $(i,j)\in\{0,\ldots,N\}^2$ we have
\begin{equation}\label{asympt}
\lim_{n\rightarrow\infty}n^{\epsilon_{ij}}(U_j^i(n)-\delta_{ij}I)=L_j^i,
\end{equation}
where $\epsilon_{ij}=\frac{1}{2}(\delta_{0i}+\delta_{0j})$. 

\smallskip
In the context of measurement, the expression (\ref{exprLi}) implies that only the asymptotic of the terms $U_j^0(n)$ are relevant. In terms of total Hamiltonian (\ref{Htot}), in \cite{FRTC} it is shown that these asymptotics for the $U^i_j$'s can be obtained by considering interaction Hamiltonian $H_I$ of type (\ref{HI}) and by considering the coupling constant $\lambda=\sqrt{n}$. In that case, they  obtain 
$$
L_k^0=-iC_k\,.
$$

Now with (\ref{asympt}), we are in position to express the main result of \cite{P3} which links discrete and continuous quantum trajectories. To this end, we introduce functions (when it has a meaning) defined on the set of states:
\begin{align*}
g_i(\rho)&=\frac{\displaystyle \sum_{k,l=1}^N\,p_{kl}^i\,L_k^0\rho(L_l^0)^\star}{\tr\left[\displaystyle\sum_{k,l=1}^N\,p_{kl}^i\,L_k^0\rho(L_l^0)^\star\right]}-\rho\\\\
v_i(\rho)&=\tr\left[\sum_{k,l=1}^N\,p_{kl}^i\,L_k^0\rho(L_l^0)^\star\right]\\
h_i(\rho)&=\frac{1}{\sqrt{p_{00}^i}}\left[\sum_{k=1}^N\big(p_{k0}^i\,L_k^0\rho+p^i_{0k}\,\rho (L_{k}^{0})^\star\big)-\tr\left[\sum_{k=1}^N\big(p_{k0}^i\,L_k^0\rho+p^i_{0k}\,\rho (L_{k}^{0})^\star\big)\right]\rho\right]\\
\mathcal{L}(\rho)&=L_0^0\rho+\rho(L_0^0)^\star+\sum_{k=1}^NL_k^0\,\rho\,(L_k^0)^\star.
\end{align*}

\begin{thm}\label{conzerotemp}
Let $A=\sum_{i=0}^p\lambda_iP_i$ be an observable of $\rH$. As $\sum_iP_i=I$, without restriction, we can assume that $p_{00}^0\neq0$. Let $I=\{i\in\{1,\ldots,p\}/p_{00}^i=0\}$ and $J=\{1,\ldots,p\}\setminus I$. Let $\rho_0$ be a state on $\mathcal{H}_0$ and let $(\rho_n(t))$ be the stochastic process defined from the discrete quantum trajectory $(\rho_k)$ by $\rho_n(t)=\rho_{[nt]}$. We then have the following convergence result.
\begin{itemize}
\item If $J=\emptyset$, the process $(\rho_n(t))$ converges in distribution to the solution of the stochastic differential equation
\begin{equation}
\rho_t=\rho_0+\int_0^t\mathcal{L}(\rho_{s-})\, ds+\sum_{i=1}^p\int_0^t\int_{\mathbb{R}}g_i(\rho_{s-})\mathbf{1}_{0<x<v_i(\rho_{s-})}\, \big[N_i(dx,ds)-dx\,ds\big],
\end{equation}
where $(N_i)_{1\leq i\leq N}$ are $N$ independent Poisson processes on $\mathbb{R}^2$.
\item If $J\neq\emptyset$, the process $(\rho_n(t))$ converges in distribution to the solution of the stochastic differential equation
\begin{eqnarray}\label{jsto}
\rho_t&=&\rho_0+\int_0^t\mathcal{L}(\rho_{s-})\,ds+\sum_{i\in J\bigcup\{0\}}\int_0^th_i(\rho_{s-})dW_i(s)\nonumber\\
&&+\sum_{i\in I}\int_0^t\int_{\mathbb{R}}g_i(\rho_{s-})\mathbf{1}_{0<x<v_i(\rho_{s-})}\,\big[N_i(dx,ds)-dx\,ds\big]
\end{eqnarray}
where $(W_i(t))_{0\leq i\leq N}$ are $N+1$ independent Brownian motions independent of the Poisson processes $(N_i)_{1\leq i\leq N}.$
\end{itemize}
\end{thm}

\subsection{The 2-Dimensional Case}\label{20}

In order to illustrate this theorem, we investigate the case where $\mathcal{H}=\mathbb{C}^2$. In this situation, we get two different behaviours depending on the fact that $p^0_{00}=1$ or not.

\smallskip
Indeed, in the case $p_{00}^0=1$ we have $J=\emptyset$ and the case $p^0_{00}\neq1$ corresponds to  $J\neq\emptyset$. Furthermore the case $p_{00}^0=1$ corresponds to a case where the observable $A$ is diagonal in the basis $\{X_0,X_1\}$, that is, of the form 
$$
A=\lambda_0\, a_0^0+\lambda_1\, a_1^1\,.
$$ 
The limit equation is then
\begin{multline}
\rho_t=\rho_0+\int_0^t\mathcal{L}(\rho_{s-})\, ds+\int_0^t\int_{\mathbb{R}}\Bigg(\frac{L_1^0\,\,\rho_{s-}\,\,(L_1^0)^\star}{\tr[L_1^0\rho_{s-}(L_1^0)^\star]}-\rho_{s-}\Bigg)\times\hfill\\
\hfill\qquad\times \mathbf{1}_{0<x<\tr[L_1^0\rho_{s-}(L_1^0)^\star]}\,\big[N(dx,ds)-dx\, ds\big]\,.\label{jump}
\end{multline}
By putting $L_1^0=C$ and by considering the process $\tilde{N}_t=\int_0^t\int_{\mathbb{R}}
\mathbf{1}_{0<x<\tr[L_1^0\rho_{s-}(L_1^0)^\star]}N(dx,ds)$, we obtain the jump equation  (\ref{J}) mentioned in Introduction. Indeed,
the process $(\tilde{N}_t)$ is a counting process with stochastic intensity $\int_0^t\tr[L_1^0\rho_{s-}(L_1^0)^\star]\, ds$. Actually,
the expression (\ref{jump}) is a rigorous way to consider jump stochastic Schr\"odinger equations (see \cite{P2}).

\smallskip
The other case $p_{00}^0\neq1$ gives rise to a diffusive equation. For example, consider the case $p_{00}^0=1/2$ (the other situations are similar). The observable $A$ has then to be of the form 
$$
A=\frac{\lambda_0}2\,(a_0^0+a_1^0+a_0^1+a_1^1)+\frac{\lambda_1}2\,(a_0^0-a_1^0-a_0^1+a_1^1)\,.
$$
Hence, we get the limit equation
\begin{eqnarray}\label{diff}
\rho_t&=&\rho_0+\int_0^t\mathcal{L}_0(\rho_s)\,ds+\int_0^t\Big(L_1^0\rho_s+\rho_s(L_1^0)^\star-\tr[L_1^0\rho_s+\rho_s(L_1^0)^\star]\rho_s\Big)\frac{\sqrt{2}}{2}\,dW_1(s)\nonumber\\
&&+\int_0^t\Big(L_1^0\rho_s+\rho_s(L_1^0)^\star-\tr[L_1^0\rho_s+\rho_s(L_1^0)^\star]\rho_s\Big)\frac{-\sqrt{2}}{2}\,dW_2(s)\,.
\end{eqnarray}
Note that by defining a Brownian motion $W_t=(\sqrt{2}/2)\, W_1(t)-(\sqrt{2}/2)\, W_2(t)$, we recover the  diffusive equation $(\ref{D})$.

In \cite{P1,P2}, the equations $(\ref{jump},\ref{diff})$ are studied in details and the convergence from discrete to continuous trajectories is obtained (with different techniques than the more general result \cite{P3}).

\section{From Discrete to Continuous Quantum Trajectories at Positive Temperature}\label{c}

All the results mentioned in previous section are based on the construction of \cite{FRTC} which makes heavy use of the fact that the reference state of $\rH$ is a pure state. Indeed, this condition is strongly needed in order to define the countable tensor product $\otimes_{n\in\NN^*} \rH$ and its continuous limit, the continuous tensor product $\otimes_{t\in\Rp} \rH$. 

When considering that the environment is made of a chain of systems $\rH$ each of which in thermal equilibrium state 
$$
\rho_\beta=\frac{1}{Z_\beta} e^{-\beta H_R}
$$
we cannot directly apply their results. The idea here follows the one developed in \cite{MR2323437}, that is, we take the G.N.S. representation of the state $\rho_\beta$. This way, the state $\rho_\beta$ becomes a pure state, but on a larger state space.

\subsection{The G.N.S Representation of the Heat Bath}\label{GNS}

The G.N.S representation of $(\mathcal{H},\rho_\beta)$, also called cyclic representation, is described as follows. At positive temperature, since the state $\rho_\beta$ is faithfull, it defines a scalar product on $\mathcal{H}'=\mathcal{B}(\mathcal{H})$ by
\begin{equation}\label{prodscal}\langle A,B\rangle=\tr\big[\rho_\beta\, A^\star B\big],\end{equation}
for all $(A,B)\in \mathcal{H}'$. 

For all $A\in\mathcal{H}',$ we denote by $\pi(A)$ the linear map from $\mathcal{H}'$ to $\mathcal{H}'$ defined by
$$\pi(A)B=AB,$$
for all $B\in\mathcal{H}'$. The linear map $\pi$ from $\mathcal{H}'$ into $\mathcal{B}(\mathcal{H}')$ is a  representation of $(\mathcal{H},\rho_\beta)$, the so-called ``G.N.S. representation". In particular we have
$$\langle I,\pi(A)I\rangle=\tr\big[\rho_\beta A\big],$$
for all $A\in\mathcal{H}'$. This means that transported by $\pi$ the action of the state $\rho_\beta$ on a observable $A$ is the same that the one of the pure state $\vert I\rangle\langle I\vert$ on $\pi(A)$. 

\smallskip
In order to compute the matrix coefficients of the operator $\pi(U)$, we need to specify an orthonormal basis of $\mathcal{H}'$. The only restriction on this basis is that it has to contain the reference state $\vert I\rangle\langle I\vert$. 

The Hilbert space $\mathcal{H}'$ is a $(N+1)^2$ dimensional space. We denote by $X_0^0$ the identity operator. Next for $i=1,\ldots,N$, we denote by $X_i^i$ the diagonal matrix with diagonal elements $\{\nu_i^0,\ldots,\nu_i^N\}$ such that
$$\langle X_i^i,X_j^j\rangle=\delta_{ij}\,,$$
for all $i,j=0,\ldots,N$. Such operators can be constructed by extending the vector $(1,\ldots,1)$ into an orthonormal basis of $\mathbb{C}^{N+1}$ for the scalar product 
$$\sum_{i=0}^N\beta_i\, \overline{x}_iy_i\,.$$
In order to complete the basis, we define $X_j^i$ for $i\neq j\in\{0,\ldots,N\}$ by
$$X_j^i=\frac{1}{\sqrt{\beta_i}}\, a_j^i\,.$$
Thus, we have construct an orthonormal basis $\{X_j^i,i,j=0,\ldots,N\}$ of $\mathcal{H}'$ for the scalar product (\ref{prodscal}).

In this basis an operator $K$ on $\mathcal{H}_0\otimes\mathcal{H}$ is transported by $\pi$ as an operator $\pi(K)=(K_{kl}^{ij})_{0\leq i,j,k,l\leq N}$ where the coefficients $K_{kl}^{ij}$ are operator on $\mathcal{H}_0$. These coefficients are given by
\begin{equation}\label{Kijkl}
K_{kl}^{ij}=\tr_{\rH}\big[(I\otimes\rho_\beta)(I\otimes X_l^k)^\star K(I\otimes X_j^i)\big]\,.
\end{equation}

\subsection{Asymptotics of $U$ in the G.N.S. Representation}

Recall that we have defined the basic unitary interaction $U$ as
\begin{equation}\label{unit}
U=exp\left(-i\frac{1}{n}\big(H_0\otimes I+I\otimes H_R+\sqrt{n}\,H_I\big)\right)\,.\end{equation}
We are in position to describe $\pi(U)$ in the asymptotic way. Here, the translation of condition (\ref{asympt}) is the existence of operators $L_{kl}^{ij}$ such that
\begin{equation}\label{UGNS}
 \lim_{n\rightarrow\infty}n^{\epsilon_{kl}^{ij}}\left(U_{kl}^{ij}-\delta_{(i,j),(k,l)}I\right)=L_{kl}^{ij}\,,
 \end{equation}
where $\epsilon_{00}^{00}=1,\epsilon_{kl}^{00}=\epsilon_{00}^{kl}=1/2$ and the others are equal to zero.

\smallskip
 We need to check that the unitary operator (\ref{unit}) provides good asymptotic. Actually in our context of indirect quantum measurement, according to Theorem \ref{conzerotemp}, we only need the expression of $L_{kl}^{00}$.

\begin{pr}\label{Lookl}
The expression of the final relevant limit operators $L_{kl}^{00}$ is given by
\begin{eqnarray}\label{exprlim}
L_{00}^{00}&=&iH_0+\sum_{i=1}^N\beta_i\gamma_iI+\frac{1}{2}\sum_{i=0}^N\big(\beta_0C_i^\star C_i+\beta_iC_iC_i^\star\big),\nonumber\\
L_{k0}^{00}&=&-i\sqrt{\beta_k}C_k^\star,
\,\,\,\,\,\,\,\,\,\,\,\,L_{0l}^{00}\,=\,-i\sqrt{\beta_0}C_l
\end{eqnarray}
\end{pr}
\begin{pf}
 In block form we have
$$
H_{\rm tot}=\left(\begin{matrix}H_0&\sqrt{n}\,C_1^\star&\sqrt{n}\,C_2^\star&\ldots&\sqrt{n}\,C_N^\star\\\\
\sqrt{n}\,C_1&H_0+\gamma_1I&0&\ldots&0\\\\
\sqrt{n}\,C_2&0&H_0+\gamma_2I&\ldots&0\\\\
\vdots&\vdots&\ldots&\ddots&\vdots\\\\
\sqrt{n}\,C_N&0&0&\ldots&H_0+\gamma_NI
\end{matrix}\right)$$
and hence the operator $U(n)$ can be shown to be of the form (cf \cite{FRTC})
$$\left(\begin{matrix}I-\frac{1}{n}iH_0-i\frac{1}{n}\gamma_0I&-i\frac{1}{\sqrt{n}}C_1^\star+\circ\left(\frac{1}{n^{3/2}}\right)&\ldots&-i\frac{1}{\sqrt{n}}C_1^\star+\circ\left(\frac{1}{n^{3/2}}\right)\\-\frac{1}{2}\frac{1}{n}\sum_{i=1}^NC_i^\star C_i+\circ\left(\frac{1}{n^2}\right)&&&\\\\
-i\frac{1}{\sqrt{n}}C_1+\circ\left(\frac{1}{n^{3/2}}\right)&I-\frac{1}{n}iH_0-i\frac{1}{n}\gamma_1I&\ldots&-\frac{1}{2}\frac{1}{n}C_1C_N^\star\\&-\frac{1}{2}\frac{1}{n}C_1 C_1^\star+\circ\left(\frac{1}{n^2}\right)&&\\\\
\vdots&\vdots&\ddots&\vdots\\\\
-i\frac{1}{\sqrt{n}}C_N+\circ\left(\frac{1}{n^{3/2}}\right)&-\frac{1}{2}\frac{1}{n}C_NC_1^\star&\ldots&I-\frac{1}{n}iH_0-i\frac{1}{n}\gamma_NI\\&&&-\frac{1}{2}\frac{1}{n}C_N C_N^\star+\circ\left(\frac{1}{n^2}\right)\end{matrix}\right).$$

Now we can compute the asymptotic form of $U_{kl}^{00}(n)$. Keeping in mind that in the appropriate basis $\tilde{B}$, the partial trace of an operator is the operator obtained by summing the diagonal blocks, we get
\begin{eqnarray}\label{oooo}
U_{00}^{00}(n)&=&\tr_{\rH}[I\otimes\rho_\beta\,X_0^0UX_0^0]=\tr_{\rH}[I\otimes\rho_\beta\,U]\nonumber\\
&=&\beta_0\left(I-\frac{1}{n}\Bigg(iH_0+\frac{1}{2}\sum_{i=1}^NC_i^\star C_i\Bigg)\right)\nonumber\\&&+\sum_{i=1}^N\beta_k\left(I-\frac{1}{n}\left(iH_0+\gamma_iI+\frac{1}{2}C_iC_i^\star\right)\right)
+\circ\left(\frac{1}{n}\right)\nonumber\\
&=&I-\frac{1}{n}\left(iH_0+\sum_{i=1}^N\beta_i\gamma_iI+\frac{1}{2}\sum_{i=0}^N\big(\beta_0C_i^\star C_i+\beta_iC_iC_i^\star\big)\right)+\circ\left(\frac{1}{n}\right)\,.
\end{eqnarray}
Let us stress that we have used $\sum_{i=0}^N\beta_i=1$ to get the expression (\ref{oooo}). Now if $k=0$ and $l\neq0$, we have
\begin{eqnarray}\label{oool}
U_{0,l}^{00}&=&\frac{1}{\sqrt{\beta_l}}\tr_{\rH}\big[(I\otimes\rho_\beta)\,\,(I\otimes a_0^l)\,U\big]\nonumber\\
&=&-i\frac{1}{\sqrt{n}}\sqrt{\beta_0}C_l+\circ\left(\frac{1}{\sqrt{n}}\right).
\end{eqnarray}
In the same way if $l=0$ and $k\neq0$, we get
\begin{equation}\label{oook}
U_{k,0}^{00}=-i\frac{1}{\sqrt{n}}\sqrt{\beta_k}C_k^\star+\circ\left(\frac{1}{\sqrt{n}}\right).\end{equation}
Actually the terms (\ref{oooo}, \ref{oool}, \ref{oook}) are the only terms which remain when considering the limit by applying (\ref{UGNS}). Indeed the other terms are expressed as
$$U_{kk}^{00}=-\frac{1}{n}\Bigg(\sum_{i=1}^N\beta_i\overline{\nu_k^i}(\gamma I+\frac{1}{2}C_iC_i^\star)+\beta_0\overline{\nu_k^0}\left(i\gamma_0I+\frac{1}{2}\sum_{i=1}^NC_iC_i^\star\right)\Bigg)+\circ\left(\frac{1}{n}\right)$$
and if $k\neq0$, $l\neq0$ and $k\neq l$
$$U_{kl}^{00}=-\frac{1}{n}\frac{1}{2}\sqrt{\beta_k}C_lC_k^\star+\circ\left(\frac{1}{n}\right).$$
Hence, by (\ref{UGNS}), they do not contribute in the limit. 
\end{pf}

\subsection{Quantum Trajectories at Positive Temperature}\label{QTPT}

Before stating the equivalent of Theorem \ref{conzerotemp} with positive temperature,  we need to be clear on how observables are transformed by the G.N.S representation. In particular, we have to describe $\pi(I\otimes P)$ when $P$ is a projector. By the rule (\ref{Kijkl}), the coefficients $P_{kl}^{ij}$ of $\pi(I\otimes P)$ are given by
\begin{eqnarray}\label{proj}
P_{kl}^{ij}&=&\tr_{\rH}\big[I\otimes\rho_\beta\,(I\otimes X_l^k)^\star\,I\otimes P (I\otimes X_j^i)\big]\nonumber\\
&=&\tr_{\rH}\big[I\otimes(\rho_\beta\,\big(X_l^k)^\star\,P\,X_j^i\big)\big]\nonumber\\
&=&\tr\big[\rho_\beta\,\big(X_l^k)^\star\,P\,X_j^i\big]\,I\,.
\end{eqnarray}
Define $p_{kl}^{ij}=\tr\big[\rho_\beta\,\big(X_l^k)^\star\,P\,X_j^i\big]$ and $P'=(p_{kl}^{ij})$. Equation (\ref{proj}) means
\begin{equation}\label{proj'}
\pi(P)=I\otimes P'\,.
\end{equation}
We are then in a similar situation as for zero temperature, but one has to notice a very important fact: the first coefficient $p_{00}^{00}$ is now always strictly positive. Indeed, we have
\begin{eqnarray}\label{positivity}
p_{00}^{00}=\tr\big[\rho_\beta\,\big(X_0^0)^\star\,P\,X_0^0\big]=\tr\big[\rho_\beta\, P\big]=\tr\big[\rho_\beta\, P^\star P\big]=\langle P,P\rangle>0\,.
\end{eqnarray} 
This simple remark has an important consequence:  {\bf there will be no jump contribution in the stochastic master equation for a heat bath}. 

Indeed, let us consider an observable $A=\sum_{i=0}^p\lambda_iP_i$ and $\pi(A)=\sum_{i=0}^p\lambda_i\,I\otimes P_i'$. In Theorem \ref{conzerotemp}, the jump contribution is directly connected to the set $I=\{i\in\{1,\ldots,p\}/p_{00}^i=0\}$. In positive temperature, with the notation $P'_m=(p_{kl}^{ij}(m))$ the analogue of set $I$ is the set $I'=\{i\in\{1,\ldots,p\}/p_{00}^{00}(i)=0\}$. Hence, the property (\ref{positivity}) implies that $I'=\emptyset$, which implies that there is no jump contribution.

\smallskip
Consider the following functions defined on the set of the states:
 \begin{eqnarray}
 \tilde{h}_i(\rho)&=&\frac{1}{\sqrt{p_{00}^{00}(i)}}\Bigg[\sum_{k=1}^N\Big(p_{k0}^{00}(i)\,L_{k0}^{00}\rho+p_{0k}^{00}(i)\,L_{0k}^{00}\rho+p_{00}^{k0}(i)\,\rho(L_{k0}^{00})^\star+p_{00}^{0k}(i)\,\rho(L_{0k}^{00})^\star\Big)\nonumber\\&&-\tr\Big[p_{k0}^{00}(i)\,L_{k0}^{00}\rho+p_{0k}^{00}(i)\,L_{0k}^{00}\rho+p_{00}^{k0}(i)\,\rho(L_{k0}^{00})^\star+p_{00}^{0k}(i)\,\rho(L_{0k}^{00})^\star\Big]\,\rho\Bigg]\,,\\
\tilde{\mathcal{L}}(\rho)&=&L_{00}^{00}\rho+\rho L_{00}^{00}+\sum_{k=1}^{N}\Big(L_{k0}^{00}\rho+\rho (L_{k0}^{00})^\star+L_{0k}^{00}\rho+\rho (L_{0k}^{00})^\star\Big)\,.\end{eqnarray}

\begin{thm}
Let $A=\sum_{i=0}^p\lambda_iP_i$ be an observable. Let $\rho_0$ be a state on $\mathcal{H}_0$. Let $(\rho_k)$ be the discrete quantum
trajectory describing the quantum repeated measurements at positive temperature. Let $(\rho_n(t))$ be the sequence of
stochastic processes defined for all $t$ and all $n$ by $\rho_n(t)=\rho_{[nt]}(t)$. Then $(\rho_n(t))$ converges in distribution,  when
$n$ goes to infinity, to the solution of the stochastic differential equation
\begin{equation}\label{diiffheat}
\rho_t=\rho_0+\int_0^t\tilde{\mathcal{L}}(\rho_s)\,ds+\sum_{i=0}^p\int_0^t\tilde{h}_i(\rho_s)\, dW_i(s)\,,
\end{equation}
where $(W_i(t))$ are $N+1$ independent Brownian motions.
\end{thm}
\begin{pf}
With the expression (\ref{proj'}), with the fact that the state $\rho_\beta$ is a pure state in the G.N.S. representation, we can apply directly Theorem \ref{conzerotemp}, with the particular restriction we have mentioned above. This gives easily Equation (\ref{diiffheat}). 
\end{pf}

\bigskip
 With the explicit expression of the coefficients $L_{kl}^{00}$, Equation (\ref{diiffheat}) can be made more explicit:
\begin{multline}\label{eqtotal}
\rho_t=\rho_0+\int_0^t\Bigg(-i[H_0,\rho_s]-\frac{1}{2}\sum_{k=0}^N\Big(\beta_0(C_k^\star C_k\,\rho_s+\rho_s\,C_k^\star C_k-2C_k\,\rho_s\,C_k^\star)\Big)\hfill\\
\hfill-\frac{1}{2}\sum_{k=0}^N\Big(\beta_k(C_kC_k^\star\,\rho_s +\rho_s\, C_kC_k^\star-2C_k^\star\,\rho_s\,C_k)\Big)\Bigg)\,ds\\
\hfill-\sum_{m=0}^p\int_0^t\frac{1}{\sqrt{p_{00}^{00}(m)}}\Bigg[\sum_{k=1}^N i\sqrt{\beta_0}\Big(p_{0k}^{00}(m)C_k\rho_s-p_{00}^{0k}(m)\rho_sC_k^\star-\tr\Big[p_{0k}^{00}(m)C_k\rho_s-p_{00}^{0k}(m)\rho_sC_k^\star\big]\rho_s\\
\hfill+\sum_{k=1}^Ni\sqrt{\beta_k}\Big(p_{k0}^{00}(m)C_k^\star\rho_s-p_{00}^{k0}(m)\rho_sC_k-Tr\Big[p_{k0}^{00}(m)C_k^\star\rho_s-p_{00}^{k0}(m)\rho_sC_k\big]\rho_s\Bigg]dW_m(s)\,.
\end{multline}
In this equation the expression
\begin{multline}
\tilde{\mathcal{L}}(\rho)=-i[H_0,\rho]-\frac{1}{2}\sum_{k=0}^N\Big(\beta_0(C_k^\star C_k\,\rho+\rho\, C_k^\star C_k-2C_k\,\rho\, C_k^\star)\Big)\hfill\\
\hfill-\frac{1}{2}\sum_{k=0}^N\Big(\beta_k(C_kC_k^\star\,\rho +\rho\, C_kC_k^\star-2C_k^\star\,\rho\, C_k)\Big)\end{multline}
corresponds to the usual Lindblad operator describing the evolution of a small system in contact with a heat bath at positive temperature.

\subsection{The 2-Dimensional Case}\label{2T}

As in Section \ref{20}, we want now to specialize the equation (\ref{eqtotal}) when $\mathcal{H}=\mathbb{C}^2$ and when considering particular observables.

\smallskip
The first case is when the observable is diagonal, that is $A=\lambda_0a_0^0+\lambda_1a_1^1$. In this case we have to compute $\pi(a_0^0)$ and $\pi(a_1^1)$. As we have $a_0^0+a_1^1=I$ and $\pi(I)=I$, we just have to compute $\pi(a_0^0)$. Since $p_{00}^{01}(0)=\overline{p_{01}^{00}(0)}$ and $p_{00}^{10}(0)=\overline{p_{10}^{00}(0)}$, we have only three terms to determine: $p_{00}^{00}(0)$, $p_{01}^{00}(0)$ and $p_{10}^{00}(0)$. 

We get
\begin{eqnarray}\label{prems}
p_{00}^{00}(0)&=&\tr[\rho_\beta\,a_0^0]=\beta_0\nonumber\\
p_{01}^{00}(0)&=&\tr[\rho_\beta(X_1^0)^\star a_0^0X_0^0]=\frac{1}{\sqrt{\beta_0}}\tr[\rho_\beta a_0^1a_0^0]=0\nonumber\\
p_{10}^{00}(0)&=&\tr[\rho_\beta (X_0^1)^\star a_0^0X_0^0]=\frac{1}{\sqrt{\beta_1}}\tr[\rho_\beta a_1^0a_0^0]=0\,.\end{eqnarray}
As a consequence the equation (\ref{eqtotal}) for a diagonal observable becomes
$$\rho_t=\rho_0+\int_0^t\tilde{\mathcal{L}}(\rho_s)\, ds$$
which is just the master equation for a heat bath, with no noise contribution. Hence, at positive temperature, the
repeated measurements of the observable $A$ gives rise to a deterministic limit behavior. Recall that at zero temperature, the limit
behavior was described by a jump equation. 

At positive temperature, we had already seen that the limit behavior should not involve jump
contribution, but here, in addition we have no randomness at all in the limit. 

\bigskip
Let us now consider the observable $A=\lambda_0/2(a_0^0+a_1^0+a_0^1+a_1^1)+\lambda_1/2(a_0^0-a_1^0-a_0^1+a_1^1)$. As we have $1/2(a_0^0+a_1^0+a_0^1+a_1^1)+1/2(a_0^0-a_1^0-a_0^1+a_1^1)=I$, we just need to compute $\pi\big(1/2(a_0^0+a_1^0+a_0^1+a_1^1)\big)$. With the previous computations, we only have to determine $\pi(a_1^0)$ and $\pi(a_0^1)$. In order to simplify the notations put $R=a_1^0$ and $S=a_0^1$. We have
\begin{eqnarray}\label{deuz}
R_{00}^{00}&=&R_{10}^{00}\,=\,S_{00}^{00}\,=\,S_{01}^{00}\,=\,0\nonumber\\
R_{01}^{00}&=&\sqrt{\beta_0},\,\,\,\,\,\,\,
S_{10}^{00}\,=\,\sqrt{\beta_1}\,.
\end{eqnarray}
Hence by  (\ref{prems}, \ref{deuz}), for $P_0=1/2(a_0^0+a_1^0+a_0^1+a_1^1)$, we get
\begin{eqnarray}
p_{00}^{00}(0)=\frac{1}{2},\,\,\,\,\,p_{01}^{00}(0)=\frac{\sqrt{\beta_0}}{2},\,\,\,\,\,p_{10}^{00}(0)=\frac{\sqrt{\beta_1}}{2}\,.\end{eqnarray}
The equation (\ref{eqtotal}) becomes
\begin{multline}
\rho_t=\rho_0+\int_0^t\tilde{\mathcal{L}}(\rho_s)\,ds
-\int_0^t\frac{i\sqrt{2}}{2}\Bigg(\beta_0\Big(C_1\rho_s-\rho_sC_1^\star-\tr[C_1\rho_s-\rho_sC_1^\star]\rho_s\Big)+\hfill\\
\hfill +\beta_1\Big(C_1^\star\rho_s-\rho_sC_1-\tr[C_1^\star\rho_s-\rho_s
C_1]\rho_s\Big)\Bigg)\, dW_0(s)\\
\qquad\qquad+\int_0^t\frac{i\sqrt{2}}{2}\Bigg(\beta_0\Big(C_1\rho_s-\rho_sC_1^\star-\tr[
C_1\rho_s-\rho_sC_1^\star]\rho_s\Big)+\hfill\\
\hfill+\beta_1\Big(C_1^\star\rho_s-\rho_sC_1-\tr[
C_1^\star\rho_s-\rho_sC_1]\rho_s\Big)\Bigg)\,dW_1(s)\,.
\end{multline} 
By defining a new Brownian motion $W_t=\sqrt{2}/2(W_0(t)-W_1(t))$ and
by putting $C=-iC_1$, the above equation becomes
\begin{multline}\label{diftemp}
\rho_t=\rho_0+\int_0^t\tilde{\mathcal{L}}(\rho_s)\, ds
+\int_0^t\Bigg(\beta_0\Big(C\rho_s+\rho_sC^\star-\tr[C\rho_s+\rho_sC^\star]\rho_s\Big)+\hfill\\
\hfill+\beta_1\Big(C^\star\rho_s+\rho_sC-Tr[C^\star\rho_s-\rho_sC]\rho_s\Big)\Bigg)\,dW_s\,.\end{multline}
The equation (\ref{diftemp}) is then the equivalent of diffusive equation (\ref{D}) (see Introduction) for the model with positive temperature.

\section*{Acknoledgments}

This work has been supported by the funding of the ANR "Hamiltonian and Markovian Approach of Statistical Quantum Physics". The second author want to thank Professor A. Barchielli and F. Petruccione for useful discussions.

\end{document}